\documentclass[prl,aps,showpacs,tightenlines,twocolumn]{revtex4}
\usepackage{graphicx}

\newcommand{\ds}{\displaystyle}

\begin{document}

\title{Adiabatic density perturbations and matter generation 
       from the MSSM} 

\author{Kari Enqvist$^{a,b}$, Shinta Kasuya$^b$,
        and Anupam Mazumdar$^c$}

\affiliation{
$^a$ Department of Physical Sciences,
     P. O. Box 64, FIN-00014, University of Helsinki, Finland.\\
$^b$ Helsinki Institute of Physics, P. O. Box 64,
     FIN-00014, University of Helsinki, Finland.\\
$^c$ Physics Department, McGill University,
     3600-University Road, Montreal, H3A 2T8, Canada.}

\date{November 11, 2002}

\begin{abstract}
We propose that the inflaton is coupled to ordinary matter only
gravitationally and that it decays into a completely hidden sector. 
In this scenario both baryonic and dark matter originate from the
decay of a flat direction of the minimal supersymmetric standard model
(MSSM), which is shown to generate the desired adiabatic perturbation
spectrum via the curvaton mechanism. The requirement that the energy
density along the flat direction dominates over the inflaton decay
products fixes the flat direction almost uniquely. The present
residual energy density in the hidden sector is typically shown to be
small.  
\end{abstract}

\pacs{98.80.Cq
\hspace{37mm} HIP-2002-53/TH, hep-ph/0211147}

\maketitle

Recent observations of the cosmic microwave background 
\cite{boomerang,maxima,dasi} appear to strongly support the
inflationary paradigm. The universe is flat and the perturbation
spectrum is adiabatic, as suggested by the simplest inflaton models
(see e.g. \cite{liddle00}). However, despite their many successes,
these models leave open the origin of ordinary baryonic matter. 
Indeed, there are no concrete particle physics models that would
specify how the inflaton couples to the Standard Model (SM) degrees
of freedom. Since the inflaton should be a gauge singlet field, it can
couple to quarks and leptons only indirectly via some other fields, or
by some nonrenormalizable couplings. Such couplings are always rather
ad hoc. 

Therefore the most natural assumption would appear to be that the
inflaton lies in a completely hidden sector and hence, except for
gravity, does not couple to the SM degrees of freedom at all. Under
such a circumstance, the inflaton decays entirely into lighter
particles in the hidden sector, and the SM matter cannot originate
from the inflaton energy density, as is usually assumed. In principle,
there could be some other hidden degrees of freedom which might
scatter to form SM particles. However, the scattering rate for
particles coupled gravitationally to SM remains less than the Hubble
rate until the temperature of the universe drops to 
$T\lesssim (m_{inf}/M_{\rm P})^4 M_{\rm P}\sim 300$~keV, where 
$m_{inf} \simeq 10^{13}$ GeV is the inflaton mass, and 
$M_{\rm P}\simeq 2.4\times 10^{18}$~GeV is the reduced Planck mass. 
This is a too low temperature for the Big Bang nucleosynthesis (BBN)
to be successful. At such low temperatures baryogenesis and the
generation of cold dark matter (CDM) would also be a great problem.

In this Letter we present a simple mechanism to create the ordinary
SM particles in the universe from a flat direction of MSSM along
which, in the absence of supersymmetry breaking and nonrenormalizable
operators, the scalar potential vanishes. The MSSM flat directions
have all been classified in \cite{gherghetta96}. They are made of
certain combinations of the squarks and sleptons, or the Higgs field
(for a review of cosmological aspects of MSSM flat directions, see
\cite{enqvist02}). It is thus obvious that the decay of the field
corresponding to the flat direction can give rise to the quarks and
leptons, together with CDM in the form of lightest supersymmetric
particles (LSPs). Moreover, as we will show, the same flat direction
can provide the adiabatic fluctuations seeding the large scale
structures in the universe. 

Our starting point is thus inflation which takes place in a hidden
sector. For a successful inflation, the energy density along the MSSM
flat directions should be negligible during inflation. There will
however be field fluctuations along those flat directions, powered by
the nonzero energy density of the inflaton; these will be
isocurvature in nature \cite{enqvist99,enqvist00} (see also
\cite{kawasaki01}). We may assume that within the single horizon
volume which is inflated to become the observable universe, only one
of the many flat directions is chosen. After the inflaton has decayed
to light degrees of freedom (in the hidden sector), the flat direction
energy density may nevertheless become dominant at a later time, and
as a consequence the original isocurvature fluctuations turn
adiabatic. Such a conversion of isocurvature fluctuations to adiabatic
fluctuations has been dubbed as the curvaton scenario in
\cite{lyth02}, and has been previously considered in the context of
pre-big bang model \cite{sloth02} (for recent discussions on
curvaton, see \cite{others}). 

The crucial requirement for an MSSM flat direction to act as a
curvaton is that it does not receive a Hubble-induced mass term during
inflation. This is known to hold true at least in D-term inflation
\cite{binetruy96} and generically in no-scale supergravity
\cite{copeland94}. In the latter case an one-loop contribution
eventually gives a Hubble-induced mass correction to the flat
directions (other than stops) of order $10^{-1}H$ \cite{gaillard95},
where $H$ is the Hubble rate during inflation. No-scale models can be
motivated, for example, by the $E_8 \times E_8'$ heterotic string
theory, where the MSSM sector resides in $E_8$ while the inflaton
sector could be found in $E_8'$. There might be other possibilities,
too. Let us here just assume that no mass term is induced by the
hidden sector inflaton $I$ and that the energy scale of the inflaton
is $V_I = \xi^4$. After inflation the hidden sector inflaton starts to
oscillate, eventually decaying into light hidden degrees of freedom
which are coupled to MSSM only gravitationally. As a consequence,
there will be no thermalization between the light hidden degrees of
freedom and the MSSM degrees of freedom, although in general there
will be a nonzero temperature in the hidden sector.

The MSSM flat direction is described by a scalar field whose potential
vanishes along that direction. However, the flatness will be lifted by
supersymmetry breaking. It will also be lifted by nonrenormalizable
terms of the form $W=\lambda\Phi^n/nM^{n-3}$
\cite{dine96,gherghetta96}, where $M$ is a cutoff scale and 
$n\gtrsim 4$ is the dimensionality of the nonrenormalizable operator;
for each flat direction, there exists a set of allowed
nonrenormalizable operators. In general, the flat direction potential
can be written as
\footnote{
Thermal corrections due to hidden radiation are negligible, since the
flat direction couples to it only gravitationally. On the other hand,
after the flat direction begins its oscillation, ordinary radiation is
gradually produced by the decay of the flat direction, but its energy
density is too small to affect the dynamics of the flat direction.}
\begin{equation}
\label{pot}
    V(\phi) = \frac{1}{2}m_{\phi}^2\phi^2
    + \frac{\lambda^2 \phi^{2(n-1)}}{2^{n-1}M^{2(n-3)}},
\end{equation}
where $\Phi=\phi e^{i \theta}/\sqrt{2}$ and the first term comes from
supersymmetry breaking so that $m_{\phi}\sim$ TeV.

After inflation, the amplitude of the flat direction is very large,
and $\phi$ will start to oscillate, eventually decaying. If the
effective mass of the decay product is heavier than the mass of the
flat direction, $g \phi>m_{\phi}$, where $g$ is some gauge or Yukawa
coupling, it can decay into light particles only through loop diagrams
involving heavy particles. On the other hand, the flat direction will
decay at tree level if $f\phi < m_{\phi}$, where $f$ is some
gauge or Yukawa coupling. Thus the decay rate is give by
\begin{equation}
    \label{decay}
    \Gamma_{\phi} \sim \left\{
      \begin{array}{ll}
          \ds{\frac{g^4 m_{\phi}^3}{\phi^2}}, & (g\phi>m_{\phi}), \\
          [3mm]
          \ds{\frac{f^2 m_{\phi}}{8\pi}}, & (f\phi<m_{\phi}).
      \end{array}
      \right.
\end{equation}
Since the flat direction is assumed to dominate the energy density
of the universe when it decays so that $H\sim m_{\phi}\phi/M_{\rm P}$,
the amplitude of the flat direction at that time is given by
\begin{equation}
    \phi_d \sim \left\{
      \begin{array}{ll}
          (g^4m_{\phi}^2M_{\rm P})^{1/3}, & (g>5\times 10^{-2}), \\
          [3mm]
          f^2 m_{\phi}, & (f<10^{-5}),
      \end{array}
      \right.
\end{equation}
The Hubble parameter at the time of decay reads as
\begin{equation}
    H_d \sim \left\{
      \begin{array}{ll}
          \ds{g^{4/3}
          \left(\frac{m_{\phi}}{M_{\rm P}}\right)^{2/3}m_{\phi}},
          & (g>5\times 10^{-2}), \\[3mm]
          f^2 m_{\phi}, & (f<10^{-5}).
      \end{array}
      \right.
\end{equation}
Numerically, these two are almost the same:
$H_d \lesssim 10^{-10} m_{\phi}$. This roughly yields a temperature of
order $10^5$~GeV, hence giving rise to a hot Big Bang universe in
the SM sector. This observable sector reheat temperature is high
enough for baryogenesis to take place at the electroweak phase
transition \cite{rubakov96} as well as for the LSPs to have the right
(thermal) abundance required for CDM
\cite{jungman96}. 

Let us now follow the dynamics of the flat direction during and
after inflation until its decay. During inflation, the energy is
dominated by the inflaton, and the flat direction quickly gets into
the slow roll regime. The amplitude during inflation is given by
\begin{equation}
    \label{phi-sr}
    \phi=\phi_{sr} \sim
    \left(\frac{\xi^2 M^{n-3}}{\lambda M_{\rm P}}\right)^{1/(n-2)}~.
\end{equation}
The amplitude of the (isocurvature) fluctuation of the flat direction,
which will be converted to the adiabatic fluctuation at a later
time, is expected to be
$P_{\zeta}^{1/2} \sim H_{inf}/\phi_{sr} \sim 10^{-5}$,
so that the scale of inflation in the hidden sector should be
\begin{equation}
    \label{xi}
    \xi^4 \sim \left( P_{\zeta}^{n-2} \lambda^{-2} \right)^{1/(n-3)} 
             \left(\frac{M}{M_{\rm P}}\right)^2 M_{\rm P}^4.
\end{equation}
Since the adiabatic fluctuation from the inflaton field is assumed to
be a subdominant component in our scenario, the energy scale of
inflation is observationally restricted by
$\xi \ll 3 \times 10^{-2} \epsilon^{1/4} M_{\rm P}$ \cite{lyth02},
where $\epsilon=(V_I'(I)/V_I)^2M_{\rm P}^2/2$ is the slow roll
parameter \cite{liddle00}, and the prime denotes a derivative with
respect to the inflaton field $I$.

During inflation the inflaton energy dominates the universe so that
the ratio
\begin{equation}
    \label{ratio-V}
    \left. \frac{V_{\phi}}{V_I} \right|_{\phi_{sr}}
           \sim \left[ \lambda^{-1} P_{\zeta}^{1/2}
           \left(\frac{M}{M_{\rm P}}\right)^{n-3}
           \right]^{2/(n-3)},
\end{equation}
will be much less than unity.

\begin{figure}
\includegraphics[width=80mm]{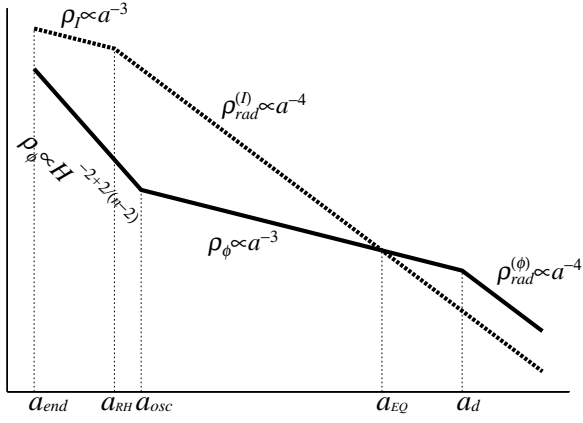}
\caption{\label{fig}
Evolution of the energy densities in the observable (solid line) and
hidden (dotted line) sectors. The subscripts `end', `RH', `osc', `EQ',
and `d' refer to, respectively, the end of inflation, reheating by
the inflaton decay creating hidden radiation, the beginning of the
flat direction oscillation in $m_{\phi}^2 \phi^2$ potential, the
equality of the energy densities of $\phi$-oscillation and hidden
radiation, and the decay of the flat direction to produce ordinary
radiation.}
\end{figure}

The evolution of the energy densities of both inflaton and the flat
direction is shown schematically in Fig.~\ref{fig}. After inflation,
the inflaton field oscillates around the minimum of its potential and
its energy density decreases as $a(t)^{-3}$, where $a(t)$ is the scale
factor of the universe. After its decay into light particles in the
hidden sector, the energy density will decrease as $a(t)^{-4}$.

On the other hand, at first the flat direction is dominated by a
nonrenormalizable term, and the field follows the boundary of a slow
roll regime, $V_{\phi}''(\phi) \sim H$, until the supersymmetry
breaking mass term begins to dominate the potential when
$H \sim m_{\phi}\approx {\cal O}(1)$~TeV. The energy density of the
inflaton is initially larger than that of the flat direction, as it
should be, so the latter never dominates the universe at this stage,
since it behaves as $\rho\propto H^{2+2/(n-2)}$, whereas 
$\rho \propto H^2$ for both the oscillating inflaton and hidden
radiation.

From then on, the flat direction field starts oscillating in a
quadratic potential with the initial amplitude
$\phi_{osc} \sim (m_{\phi} M^{n-3}/\lambda)^{1/(n-2)}$, and the energy
density decreases as $a(t)^{-3}$. The field will decay when
$H \sim \Gamma_d$. For a successful scenario, the flat direction has
to dominate the universe at the time of decay (note the difference
from the usual curvaton scenarios where the curvaton does not
necessarily have to dominate the universe). One can achieve such a
situation if the Hubble rate $H_{EQ}$ at the time when 
$\rho_{\phi}=\rho_{rad}^{(I)}$ satisfies $H_{EQ} > H_d$.

In order to evaluate $H_{EQ}$, let us first assume that the flat
direction starts its oscillation after the universe is dominated by
hidden radiation ($H_{osc} < H_{RH}$). This is the case shown in
Fig.~\ref{fig}. Imposing the condition $H_{EQ} > H_d$, we get the
constraint on $\lambda$ (see Eq.(\ref{pot})) as
\begin{equation}
    \label{lambda}
    \lambda \lesssim f^{-(n-2)/2} \left(\frac{m}{M_{\rm P}}\right)
                     \left(\frac{M}{M_{\rm P}}\right)^{n-3},
\end{equation}
where we take the decay rate of the flat direction to be
$\Gamma_d \sim f^2 m/8\pi$ (see Eq.(\ref{decay})); if the decay
products are very heavy, one should replace $f$ by 
$g^{2/3}(m/M_{\rm P})^{1/3}$). 

In the opposite case, where the flat direction oscillations begin
while the inflaton oscillations still dominate the universe, i.e.,
$H_{osc} > H_{RH}$, the energy density of the flat direction will
evolve as $\rho_{\phi} \propto H^2$, while 
$\rho_{\phi} \propto H^{3/2}$ in the hidden radiation-dominated
universe. Imposing both $H_{EQ} > H_d$ and  $H_{osc} > H_{RH}$, we
obtain exactly the same constraint as given in Eq.(\ref{lambda}).

In addition, we must demand that (a) the amplitude of the flat
direction is smaller than the cutoff scale, $\phi_{sr} < M$, that (b) 
the scale of the inflation is smaller than the COBE constraint,
$\xi < 3 \times 10^{-2} \epsilon^{1/4} M_{\rm P}$, and that (c) the
energy density of the flat direction is smaller than that of the
inflaton during inflation, $V(\phi)/V_I <1$ at $\phi=\phi_{sr}$. These
conditions from Eqs.(\ref{phi-sr}), (\ref{xi}), and
(\ref{ratio-V}) can be expressed as constraints on $\lambda$, which
respectively read as
\begin{eqnarray}
    {\rm (a)} \quad \lambda & \gtrsim & P_{\zeta}^{1/2}, \\
    {\rm (b)} \quad \lambda &
                   \gtrsim & (10^3)^{n-3} P_{\zeta}^{(n-2)/2}
                   \epsilon^{-(n-3)/2}
                   \left(\frac{M}{M_{\rm P}}\right)^{n-3}, \\
    {\rm (c)} \quad \lambda & \gtrsim & P_{\zeta}^{1/2}
                        \left(\frac{M}{M_{\rm P}}\right)^{n-3}.
\end{eqnarray}

Comparing these with the upper bound estimated above in
Eq.(\ref{lambda}), we find that there are consistent scenarios only
for $n \ge 6$, since $P_{\zeta} \sim 10^{-10}$ ($n=6$ case is
marginally consistent). However, we should also take note of the
fact that the extra energy density in the universe at the
BBN time should be less than about 10\% of the total density, which
comes from the fact that the extra neutrino species (or more precisely
relativistic degrees of freedom) should contribute $\Delta N_{\nu}<1$. 
If the oscillation of the flat direction dominates the universe long
enough, say $H_{EQ} \gtrsim 10 H_d$, this condition is easily
satisfied. Therefore, we conclude that only $n \ge 7$ flat directions
are consistent with our scenario. 

Note that the thermal bath in the hidden sector is completely
decoupled from the SM sector, and hence its temperature is expected to
be much lower than the temperature in the SM sector. The hidden energy
density is given by \cite{EKM}
\begin{equation}
    \rho_{\rm hid} = \left(a_d\over a_{EQ}\right)^{-1}\rho_{\rm obs}
      \sim \left[f^2\left({\lambda M_{\rm P} 
          \over m_{\phi} }\right)^{4 \over n-2} \beta \right]^{2/3}
          \rho_{\rm obs}, 
\end{equation}
where $\beta=1$ for $H_{osc} < H_{RH}$, and $\beta=m_{\phi}/H_{RH}$
for $H_{osc} > H_{RH}$, and $M=M_{\rm P}$ is assumed, so that the
hidden radiation can constitute typically at most 10\% of the total
energy density (if the hidden sector contains massive particles whose
energy density will eventually dominate, one should compare the
density with the CDM density).

As is well known, most of the MSSM flat directions carry baryon and/or
lepton numbers, and act like the Affleck-Dine (AD) field, which
creates baryon (lepton) asymmetry by virtue of the helical motion
induced by the A-term in the potential \cite{enqvist02,dine96}. In the
present case  the energy density of the AD field dominates the
universe, and the baryon-to-entropy ratio would in general become of
order unity, which is $10^{10}$ times larger than the value predicted
by the BBN \cite{olive00}. One way out of this impass\'e is to
suppress the phase $\theta$ of the flat direction to be $10^{-10}$,
since the baryon number is proportional to 
$\sin(2\theta) \sim 2\theta$. However, in that case the amplitude of
the isocurvature fluctuation of flat direction, 
$\delta\theta/\theta \sim H_{inf}/\phi\theta$, becomes unacceptably
large. Therefore, if the flat direction is to act as a curvaton, it
cannot posses any baryon (or lepton) number. More precisely, since the
flat direction decays well above the electroweak scale, it can carry
baryon and lepton numbers, but not $B-L$. Otherwise, there will be a
huge baryonic and/or leptonic isocurvature fluctuation. Unfortunately,
no such  direction with $n \ge 7$ exists in MSSM.

However, there are two directions which do not receive contributions
to A-terms from the nonrenormalizable superpotential which lifts the
flatness. They are the directions $n=7$ $LLddd$ (lifted by
$H_uLLLddd$) and $n=9$ $QuQuQue$ (lifted by $QuQuQuH_dee$). Since they
are lifted by superpotentials of the form $\phi^{n-1}\psi/M^{n-3}$,
where $\psi$ is the field other than the flat direction,
$\langle \psi \rangle =0$ usually leads to a vanishing A-term 
\footnote{
For a nonminimal K\"{a}hler potential, it is possible to have some
A-terms. In this situation, it might be possible to have a right
amount of the baryon asymmetry from the flat direction which acts as
a curvaton \cite{EKM}.}.
Thus, these two directions do not produce any baryon and/or lepton
number, and are free of disastrous isocurvature fluctuations. They
seem to be excellent candidates for all matter and for the adiabatic
density perturbations in models where inflation takes place in a
hidden sector.

To summarize, we have discussed a simple scenario where the decay
\footnote{
Depending on the SUSY parameter values, the flat direction could first
fragment to form $Q$ balls.} 
of the $n=7$ or 9 MSSM flat direction creates ordinary matter such as
quarks, leptons, and the LSP dark matter, simultaneously providing
adiabatic density perturbations. Since the reheating temperature for
the ordinary visible radiation comes out to be $\sim 10^5$ GeV,
baryogenesis can take place at the electroweak phase transition. 
Moreover, the reheat temperature is low enough to avoid any gravitino
problem \cite{ellis84} in the visible sector \footnote{
Gravitinos could, however, be produced thermally also by the hidden
sector radiation. For low enough reheat temperature in the hidden
sector, or for large enough entropy production due to the decay of the
flat direction in the visible sector, there should be no gravitino
problem \cite{EKM}.} 
. 

Although the situation described in this Letter is rather radical in
the sense that the inflaton is now in a completely hidden sector, such
an assumption could in fact be considered very natural on the basis
that the inflaton should anyway couple extremely weakly to the SM. 
Such ``hidden inflation" has the virtue that generation of matter and
density perturbations can take place in a sector which is testable in
future accelerator experiments.

\vspace{2mm}

S.K. is grateful to M. Kawasaki for useful discussions. K.E. is
supported partly by the Academy of Finland grant no. 51433, and
A.M. is a CITA-National fellow and acknowledges Cliff Burgess,
Zurab Berezhiani and Andrew Liddle for helpful discussion.



\end{document}